\newcommand*{\ARXIV}{}%

\ifdefined\ARXIV
  \documentclass[11pt]{article}
  \pdfoutput=1
\else
  \documentclass{siamart1116}
\fi

\usepackage{amsfonts}
\usepackage{amsmath}
 \usepackage{relsize}
 \usepackage{float}
\usepackage{graphicx}
\usepackage{epstopdf}
\usepackage{algorithm}
\usepackage{float}
\usepackage{mathtools}
\usepackage{algpseudocode}
\usepackage{booktabs} 
\usepackage{bm}
\usepackage{hyperref}

\ifdefined\ARXIV
  \usepackage{geometry}
  \usepackage{hyperref}
  \newcommand{\email}[1]{\href{mailto:#1}{\texttt{#1}}}
  \usepackage{amsthm}
  \newtheorem{remark}{Remark}

  \newcommand\figref{Fig.~\ref}
\else
  \ifpdf
    \DeclareGraphicsExtensions{.eps,.pdf,.png,.jpg}
  \else
    \DeclareGraphicsExtensions{.eps}
  \fi

\usepackage[textsize=small]{todonotes}
\newcommand{\figref}[1]{\figurename~\ref{#1}}
\numberwithin{theorem}{section}

\newtheorem{remark}{Remark}

\newcommand{\TheTitle}{Fractional Operators Applied to Geophysical Electromagnetics} 
\newcommand{\TheAuthors}{C.J. Weiss, B.G. van Bloemen Waanders, and
  H. Antil}

\headers{\TheTitle}{\TheAuthors}

\title{{\TheTitle}\thanks{Submitted to the editors DATE.
\funding{This work is partially supported by Sandia National Laboratories LDRD program.
In addition, H. Antil is partially supported by NSF grants DMS-1521590 and DMS-1818772 and Air Force Office of Scientific Research under Award NO: FA9550-19-1-0036.
Sandia National Laboratories is a multimission laboratory managed and operated by National Technology and Engineering Solutions of Sandia LLC, a wholly owned subsidiary of Honeywell International, Inc., for the U.S. Department of Energy's National Nuclear Security Administration under contract DE-NA-0003525.
This paper describes objective technical results and analysis. Any subjective views or opinions that might be expressed in the paper do not necessarily represent the views of the U.S. Department of Energy or the United States Government.
}}}
\author{
 Chester J. Weiss\thanks{Geophysics Dept., Sandia National Labs,
    Albuquerque, NM 87185-0750, USA.  cjweiss@sandia.gov
    }
  \and
  Bart G. van Bloemen Waanders\thanks{Optimization and Uncertainty Quantification, Sandia National Laboratories, P.O. Box 5800, Albuquerque, NM 87185-1320,
   \email{bartv@sandia.gov}, \url{http://www.cs.sandia.gov/\~bartv/main.html} }
\and
Harbir Antil\thanks{Dept. of Mathematical Sciences, George Mason
   University, Fairfax, VA 22030, USA. }
}

\usepackage{amsopn}

  \ifpdf
  \hypersetup{
   pdftitle={\TheTitle},
   pdfauthor={\TheAuthors}
  }
  \fi

\fi

\begin{document}

\ifdefined\ARXIV
  \title{Fractional Operators Applied to Geophysical Electromagnetics}
  \author{ 
    \thanks{{\bf Funding.} This work is partially supported by Sandia National Laboratories LDRD program.
In addition, H. Antil is partially supported by NSF grants DMS-1521590 and DMS-1818772 and Air Force Office of Scientific Research under Award NO: FA9550-19-1-0036.
Sandia National Laboratories is a multimission laboratory managed and operated by National Technology and Engineering Solutions of Sandia LLC, a wholly owned subsidiary of Honeywell International, Inc., for the U.S. Department of Energy's National Nuclear Security Administration under contract DE-NA-0003525.
This paper describes objective technical results and analysis. Any subjective views or opinions that might be expressed in the paper do not necessarily represent the views of the U.S. Department of Energy or the United States Government.}
    Chester J. Weiss\thanks{Geophysics Dept., Sandia National Labs,
    Albuquerque, NM 87185-0750, USA.  cjweiss@sandia.gov
    }
  \and
  \and  
   Bart G. van Bloemen Waanders\thanks{Optimization and Uncertainty Quantification, Sandia National Laboratories, P.O. Box 5800, Albuquerque, NM 87185-1320,
   \email{bartv@sandia.gov}, \url{http://www.cs.sandia.gov/\~bartv/main.html} } 
  \and
  \and Harbir Antil\thanks{Department of Mathematical Sciences, George Mason University, Fairfax, Virginia 
    (\email{hantil@gmu.edu}, \url{http://math.gmu.edu/\~hantil/}).}    
  }
\else
\fi

\maketitle

\begin{abstract}
  A growing body of applied mathematics literature in recent years has
  focussed on the application of fractional calculus to problems of
  anomalous transport.  In these analyses, the anomalous transport (of
  charge, tracers, fluid, etc.) is presumed attributable to
  long--range correlations of material properties within an inherently
  complex, and in some cases self-similar, conducting medium. Rather
  than considering an exquisitely discretized (and computationally
  intractable) representation of the medium, the complex and spatially
  correlated heterogeneity is represented through reformulation of the
  PDE governing the relevant transport physics such that its
  coefficients are, instead, smooth but paired with fractional--order
  space derivatives.  Here we apply these concepts to the scalar
  Helmholtz equation and its use in electromagnetic interrogation of
  Earth's interior through the magnetotelluric method.  We outline a
  practical algorithm for solving the Helmholtz equation using
  spectral methods coupled with finite element discretization.
  Execution of this algorithm for the magnetotelluric problem reveals
  several interesting features observable in field data:
  long--range correlation of the predicted electromagnetic fields; a
  power--law relationship between the squared impedance amplitude and
  squared wavenumber whose slope is a function of the fractional
  exponent within the governing Helmholtz equation; and, a
  non--constant apparent resistivity spectrum whose variability arises
  solely from the fractional exponent.  In geologic settings
  characterized by self--similarity (e.g. fracture systems; thick and
  richly--textured sedimentary sequences, etc.) we posit that
  diagnostics are useful for geologic characterization of features far
  below the typical resolution limit of electromagnetic methods in
  geophysics.
\end{abstract}

\ifdefined\ARXIV
\else

\fi

\section{Introduction}

Anomalous diffusion has been at the heart of considerable research
directed at understanding non-standard, or ``anomalous'', 
transport behavior where the mean squared displacement of random 
walk particles no longer adhere to a linear relationships with time.  
As a result, such systems reveal power laws indicative of sub
and super-diffusive behavior.  Anomalous diffusion can be described
through random walks endowed by heavy tail distributions and is
captured through non-integer exponents on time and space derivatives.
Fractional derivatives in space model super-diffusion 
and are related
to long power-law particle jumps, whereas fractional derivatives on
temporal derivatives model sub-diffusion that are induced through 
long waiting times between particle jumps.  Such behavior has
been observed in many applications including reaction-diffusion,
quantum kinetics, flow through porous media, plasma transport,
magnetic fields, molecular collisions, and geophysics applications. We
refer the reader to Metzler and Klafter
\cite{Metzler_and_Klafter:2000} for a detailed description of
anomalous diffusion, including a comprehensive list of applications.
The underlying cause for anomalous behavior in these applications is
the presence of complex structures or mechanisms that either promote
sub-diffusion or super-diffusion.  For instance, in fluid flow through
porous media, complex permeability fields cause sub-diffusive transport
through trapping mechanisms or solid/liquid surface-chemistry kinetics, ultimately
inducing a memory--type effect \cite{Caputo_WRR_00,
  Deseri_CNSNS_15}.  Super-diffusive responses have been experimentally
observed in diffusion-reaction system where the variance of a chemical
wave exceeds a linear temporal relationship \cite{Kameke_PR_81}.  This 
is interpreted to be a result of non-local interactions over distances 
beyond that to the nearest neighbors.  In the diffusion-reaction case, 
vortices in a chaotic velocity field introduce flow paths that exceed 
standard diffusion rates.

A range of natural phenomena can be described as processes in which a
physical quantity is constantly undergoing small, random fluctuations.
Such Brownian motion can be interpreted as a random walk that,
according to the central limit theorem, approaches a normal
distribution as the number of steps increases. A macroscopic
manifestation of Brownian motion is defined as diffusion whereby a
collection of microscopic quantities tends to spread steadily from
regions of high concentration to regions of lower concentration.
Through Fick's first and second laws, macroscopic particle movement
can be captured by the familiar diffusion equation, the solution of
which is identical to a normal distribution and corresponds to the
random walk probability density.  These well known concepts provide
the underpinning to investigate phenomena that violate the standard
diffusive regime.  An application space which has received relatively 
little attention but is poised for further exploration is low-frequency
electromagnetic imaging and interrogation of Earth's subsurface, a
classic geophysical exploration technique  premised on diffusive 
(transport) physics, either through the mobility of solid-state defects
in crystalline materials and free electrons in metals, or electrolytic conduction in fluids
\cite{Karato_and_Wang:2013}.   

Capturing anomalous diffusion in partial differential equations poses
considerable mathematical and numerical challenges, in particular in
the area of 1) imposing non-zero boundary conditions, 2) validating
fractional behavior for different physics, and 3) achieving
computational efficiency to realize scalable performance.  To solve
the fractional Laplacian an integral or a spectral definition can be
considered.  The choice of method however remains an open question, in
particular for non-zero boundary conditions. In this paper we consider
the spectral definition and justify this choice based on the authors'
previous developments.  Validating fractional PDEs against field
observations, laboratory measurement, or analytic solutions is
difficult, in part, because fractional calculus development has been
somewhat isolated from engineering and science applications
community. In this paper, we offer results to help bridge that gap between the 
observational science and mathematics communities. In particular, fractional concepts are applied to
geophysical electromagnetics to better characterize the subsurface and
subsequently validated through field observations, as well as
geophysical insight. A key challenge in simulating fractional PDEs is
achieving computational efficiency.  In our work we pursue an approach
that leverages the Dunford--type integral representation which, in the
case of fractional Laplacians, is computationally very attractive because
of an embarrassingly parallel loop for solving multiple Laplacians.
This parallelism however does not map to our Helmholtz
implementation because of coupling cross-terms, a fact which impacts our 
eventual desire to make use of adjoint-based optimization. Nonetheless, we 
begin with Bonito and Pasciak's solution strategy \cite{ABonito_JEPasciak_2015a,
  Bonito_and_Pasciak:2017}, with its resultant Helmholtz coupling, 
and augment it with the lifting (splitting) strategy of Antil
\cite{Antil_etal:2018a} for handling non-zero boundary conditions to 
arrive at a unique approach to solving the fractional Helmholtz equation,
regardless of its particular application here in geophysical 
electromagnetics.

The aim of this paper is to explore anomalous diffusion in the context
of geophysical electromagnetics and to derive mathematical and
algorithmic strategies for practical simulation capabilities.
Subsurface geology can exhibit complex features ranging from
hierarchical structures to self-similar geometries.  Electromagnetic
energy exposed to such media will likely produce signals significantly
different from homogenous and isotropic media.  For example,
non-local electromagnetic effects have been observed in near-surface, 
geotechnical engineering settings as a result of complex 
structured conductivity in the subsurface
\cite{Everett_GRL_02,Weiss_and_Everett:2007, Ge_etal:2015}.  Other
phenomena in the subsurface 
have been demonstrated in the context of
fluid flow in porous media and would have equivalent non-local effects
on geophysical electromagnetics \cite{Caputo_WRR_00,
  Deseri_CNSNS_15,Benson-00}. Fractured systems have been reported to
be linked to non-local effects \cite{Weiss_and_Everett:2007,
  Ge_etal:2015}, as well as layered media \cite{Elser_etal:2007}.
We are interested in detecting small scale features in the geological
subsurface (e.g. fine-scale stratigraphic laminations or regions permeated 
by fractures) that aggregate into a hierarchical ``meta-material'' while also,
by practical necessity, avoiding the detailed and computationally explosive 
discretization required to represent each of them in a given numerical simulation.
Following Caputo's strategy of replacing the standard scalar
permeability and gradient combinations in Darcy's law with a
fractional derivative \cite{Caputo_WRR_00}, we replace the Ohm's
empirical relationship with a fractional derivative to arrive at a
fractional Helmholtz equation.

Our main contributions consist of 1) deriving fractional Helmholtz via
Ampere's law by introducing a fractional spatial differential operator
into Helmholtz to account for non-local conductivity; the solution
process of this fractional partial differential equation (PDE)
requires a decomposition to separate boundary conditions from the
fractional Laplacian operator; 2) implementing computationally efficient
methods through a combination of spectral characterization of the
Laplacian, finite element discretization, and a Dunford--type 
integral representation; a reformulation allows for sparse Jacobians
and a scaling adjustment provides for a much improved conditioning; 3)
validating EM behavior through magnetotullerics.
The remainder of the paper is organized by first deriving the
fractional Helmholtz equation through a fractional gradient
relationship between the magnetic field and the underlying electric
conductivity.  Next the fractional Helmholtz equation with non-zero
boundary conditions is decomposed to separate non-zero right-hand side
and non-zero boundary conditions.  The separation provides a
convenient solution strategy and leverages the Dunford--type integral
approach to numerically solve one of the remaining equations with a
fractional Laplacian.  After the mathematical formulation, our finite
element implementation is verified through the methods of manufactured
solutions.  Finally, our numerical capability is demonstrated on a
relevant magnetotullerics application.  Our numerical results are
validated through field measurements and geophysical insight.

\section{Mathematical Formulation}

The formulation that follows consists of multiple steps.  We start by
motivating the fractional derivative operator in the context of geophysical
electromagnetics and define Ohm's constitutive law in terms of a fractional 
space derivative, later moving that derivative to the Laplacian term of the Helmholtz PDE.  
Given a fractional Helmholtz
equation with non-zero boundary conditions, a two-equation
decomposition provides a grouping of boundary conditions, source
terms, and fractional operators that allow for convenient solution
strategies.  One equation with non-homogeneous boundary conditions is
transformed to non-fractional form by deriving the ``very weak form''
so that standard solution techniques can be applied.  For the
remaining equation with a fractional Laplacian and homogeneous
boundary conditions, we appeal to a spectral representation and
resolvent formalism whereby the fractional Laplacian is transformed to
a summation of standard Laplacians using a Dunford--type integral
with appropriate quadrature.  The solution to the final system of
equations is detailed in section (\ref{numerical}) in which a finite
element discretization of both equations results in a large and
dense coefficient matrix that requires further manipulation to achieve
efficient solution performance.

We start with Faraday's law in the frequency domain with the Fourier convention of 
time derivatives $\partial_t$ mapping to the frequency domain $\omega$ as 
$\partial_t\mapsto \text{i}\omega$ and assuming constant magnetic permeability 
$\mu_0=4\pi\times 10^{-7}$ H/m:
\begin{equation}
\nabla\times \bold{E} = -\text{i}\omega\mu_0\bold{H}, 
\label{faraday}
\end{equation}
relating the curl of electric field $\bold E$ to time variations in magnetic field $\bf H$.
Paired with this is Amp\`ere's Law, $\nabla\times \bold{H} = \bold{J}$,
where $\bold J$ is the total electric current density: The sum of Ohmic currents; 
Maxwell's discplacement current $\text{i}\omega\varepsilon\bold{E}$; and, any impressed
external currents due to antennas and electrodes.  
Typically, for simple linear, isotropic materials the Ohmic currents are described by
the product of electrical conductivity $\sigma$ and electric field and at suffiently
low frequencies $\sigma >> \omega \varepsilon$ Maxwell's displacement current 
can safely be ignored. 
In a similar fashion as Caputo \cite{Caputo_WRR_00} in which he replaced the permeability 
in Darcy's equation with a time--fractional derivative, we replace the simple conductivity/field 
product in Ohm's law with a space--fractional derivative.   Preserving the
symmetry of both positive and negative power law jumps in the $z$ direction for a 
two--sided stable diffusion process requires both positive and negative fractional 
derivatives \cite[p15]{Meerschaert_and_Sikorskii:2012} 
$\mathcal D^\alpha_z = {1\over 2\cos({1\over 2}\pi\alpha)}{\left ( {\partial^\alpha\over{\partial z^\alpha}} + 
{\partial^\alpha\over\partial (-z)^\alpha} \right ) }$ 
which we normalize by the factor $\cos ({1\over 2}\pi\alpha)$ to preserve magnitude invariance under
$\alpha$.    As a consequence, the space--Fourier transform for this operator maps to the wavenumber $\nu$ domain as 
$D_z^\alpha\mapsto |\nu|^\alpha$.  Note that had a one--sided derivative with Fourier mapping
$(\pm \text{i}\nu)^\alpha$ been used, the unit--magnitude prefactor $(\pm\text {i})^\alpha$ could 
be interpreted as rotating the electrical conductivity into the complex plane, effectively 
reintroducing the Maxwell displacement current
and turning the Maxwell derivation into a mixed diffusion/wave propagation problem rather than a strictly diffusive one. 
Unlike Caputo's attempt to
emulate memory effects in permeability, our hypothesis is that certain
non-local conductivity properties create superdiffusive behavior and
can be represented by a spatial fractional derivative. 
With this fractional Ohm's law in the 
low--frequency limit (and no external sources) inserted into Amp\`ere's law, the curl 
of \eqref{faraday}  is thereby:
\begin{equation}
\nabla\times\nabla\times \bold{E} = -\text{i}\omega\mu_0\mathcal{D}_z^\alpha\left [ \sigma_{\alpha,z}\bold{E}\right ],
\label{curlcurl}
\end{equation}
where $\mathcal{D}_z^\alpha$ is the $\alpha$--order fractional
derivative in the $z$ direction and $\sigma_{\alpha,z}$ is the
electrical conductivity in units of S/m${}^{1-\alpha}$. 
For an Earth model whose conductivity
varies only as a function of vertical coordinate $z$, subject to a
vertically incident electric field oriented in the horizontal $x$
direction, the electric fields in the Earth are everywhere horizontal
such that $\bold{E}=\hat x \,u(z)$ is the primary state variable that
needs further consideration.  Furthermore, for dimensional consistency
in the fractional calculus methodology described in the following
section, we non--dimensionalize with respect to the $z$ coordinate
such that $z\mapsto\zeta=z/z^*$ to arrive at
\begin{equation}
-{\text{d}^2u\over\text{d}\zeta^2}\left ( {1\over z^*}\right )^2 + \text{i}\omega\mu_0\mathcal{D}_\zeta^\alpha
\left [ \sigma_{\alpha,\zeta}\,u(\zeta) \right ]= 0
\label{fhelm}
\end{equation}
which, after action by $\mathcal{D}_\zeta^{-\alpha}$ and generalization to 3D, becomes
\begin{equation}
\left (-\Delta_\zeta\right )^s u + \text{i}\kappa^2\,u(\zeta) = 0,
\label{fhelm_kappa}
\end{equation}
where $\left ( -\Delta_\zeta \right )^s$ is the fractional--order Laplacian in dimensionless
coordinate $\zeta$, $s = 1-{1\over 2}\alpha$ and $\kappa^2$ the dimensionless squared wavenumber 
$\omega\mu_0\sigma_{\alpha,\zeta}\left ( z^*\right )^2$.  Note that in Eq \eqref{fhelm} 
the conductivity $\sigma_{\alpha,z}$ possesses fractional length dimensions to retain 
consistency with the fractional derivative operator $\mathcal{D}_z^\alpha$.  However, 
through the non-dimensionalization process transforming Eq (\ref{fhelm}) to Eq (\ref{fhelm_kappa})
we see that the conductivity $\sigma_{\alpha,\zeta}$ reclaims its familiar, integer-ordered
units of S/m, thus avoiding awkward, fractional--dimensioned 
conductivities reported elsewhere \cite{Everett:2009,Ge_etal:2015}.  

We observe that the fractional exponent is on the Laplacian term and
in combination with the Helmholtz term motivate the challenge of a
solution strategy.  An additional complication is the need to
incorporate non-trivial boundary conditions, such as special radiation
or self-absorbing boundary conditions.  We address these issues
through the use of linear decomposition, a Dunford type integral formulation
and the very weak form for finite element discretizations.
We write the generalized fractional--order Helmholtz equation as
\begin{equation}
\label{cjw1}
\begin{aligned}
\left (-\Delta\right )^{s}u -k^2 u & = f \quad \mbox{in}\,\,\Omega, \\
 u & = g  \quad \mbox{on}\,\,\Gamma,
\end{aligned}
\end{equation}
where $k^2 = -\text{i}\kappa^2$ is introduced to simplify notation for our electromagnetic problem, but in fact, transcends
this particular choice of physics, and $(-\Delta)^s$ is understood
to be the spectral fractional Laplacian operator for non--zero Dirichlet boundary conditions
\begin{equation}
\left ( -\Delta \right )^s u(\bm{x}):= \sum_{k=1}^{\infty} \left ( \lambda_k^s\int_\Omega u\,\varphi_k \,{\text d}\Omega +
\lambda_k^{s-1}\int_\Gamma u\,\partial_n\varphi_k\,{\text d}\Gamma \right ) \varphi_k(\bm{x}),
\label{spectral}
\end{equation}
where $\varphi_k$ are the eigenfunctions of the Laplacian with corresponding 
eigenvalues $\lambda_k$ \cite[Def. 2.3]{Antil_etal:2018a} and $\bm{x}\in\Omega$ is the coordinate 
of interest. Moreover, $u$ is assumed to be sufficiently smooth. 
As it is customary in the PDE theory, we have stated this definition for smooth 
functions, however by standard density arguments it immediately extends to 
Sobolev spaces, we refer to \cite{Antil_etal:2018a} for details. 
In addition, we emphasize that when $u = 0$ on the boundary $\Gamma$, the 
definition above is nothing but the standard spectral fractional Laplacian $(-\Delta_0)^s$ 
with zero boundary conditions. We will omit the subscript 0 when it is clear from the
context.

Following this earlier work on fractional Poisson equation, we extend the basic approach 
to fractional Helmholtz and break $u$ into two parts thusly:
Let $v$ solve
\begin{equation}
\label{cjw2}
\begin{aligned}
\left (-\Delta\right )^{s}v -k^2 \left (v + w\right ) & = f \quad \mbox{in}\,\,\Omega, \\
 v & = 0  \quad \mbox{on}\,\,\Gamma,
\end{aligned}
\end{equation}
and let $w$ solve
\begin{equation}
\label{cjw3}
\begin{aligned}
\left (-\Delta\right )^{s}w &= 0 \quad \mbox{in}\,\,\Omega, \\
w & = g  \quad \mbox{on}\,\,\Gamma.
\end{aligned}
\end{equation}
Summing \eqref{cjw2} and \eqref{cjw3} it is evident that $u = v + w$.
The presence of the homogeneous boundary condition on \eqref{cjw2}
allows for a spectral representation of the Dirichlet
fractional--power Laplacian operator.  Furthermore, it has been shown
\cite{Antil_etal:2018a} that solving \eqref{cjw3} is equivalent to
solving the standard, integer--power Laplacian equation in the
very--weak sense \cite[c.f.]{Berggren:2004, Lions_and_Magenes:1972},
which in the case of smooth $g$ is simply the more--familiar weak
sense.  A simple algebraic manipulation of the spectral decomposition
provides a Laplacian with integer exponents (see
Theorem 4.1 and the subsequent proof in \cite{Antil_etal:2018a} for additional
details). Hence, we may replace \eqref{cjw3} with the following:
\begin{equation}
\label{cjw4}
\begin{aligned}
\left (-\Delta\right )w &= 0 \quad \mbox{in}\,\,\Omega, \\
 w & = g  \quad \mbox{on}\,\,\Gamma,
\end{aligned}
\end{equation}
which, when solved simultaneously with \eqref{cjw2}, yields the solution
to the original equation \eqref{cjw1}.

To solve equation (\ref{cjw2}), we follow others
\cite[e.g.]{ABonito_JEPasciak_2015a, antilshort} in
using spectral analysis of linear operators and resolvent formalism.
Specifically, we start with Kato's definition of fractional
powers for linear operators (\cite{Kato_60} Theorem 2 and supporting proof) and
simplify by setting Kato's $\lambda$ coefficient to zero. 
{This definition due to Kato coincides with \eqref{spectral} when the function values are 
zero on the boundary, as in this case the surface integral over $\Gamma$ 
vanishes which is indeed the case in \eqref{cjw2}. The Kato's definition
after applying a variable 
transformation that results in a symmetric integral, which is approximated
through quadrature is:
 
\begin{equation}
\label{cjw5}
\begin{aligned}
\left (-\Delta\right )^{-s} &= {\sin s\pi \over \pi} \int_{-\infty}^\infty
e^{(1-s)y} \left ( e^y - \Delta \right )^{-1}\,{\text d}y, \\
& \approx {\sin s\pi \over \pi} m 
\sum_{\ell=-N^-}^{N^+} 
e^{(1-s)y_\ell} \left ( e^{y_\ell} - \Delta \right )^{-1},
\end{aligned}
\end{equation}
where the quadrature nodes are distributed uniformly as $y_l = m\ell$.
Accuracy of the quadrature representation of this continuous integral
is a function of the constants $m$, $N^-$ and $N^+$ and has been shown
to be exponentially convergent \cite{ABonito_JEPasciak_2015a}.  The
constants are chosen such that the quadrature error is balanced with
the error in the spatial discretization of \eqref{cjw2}
\cite[c.f. when $w=0$]{ABonito_JEPasciak_2015a}.  In the case of a finite element
solution with linear nodal basis functions on the unit interval and
node spacing $h$, they are
\begin{equation}
\label{sinc_constants}
m = {1\over \ln{1\over h}} , \quad
N^+ = \Big\lceil{\pi^2\over 4s\,m^2}\Big\rceil , \quad\text{and}\quad
N^- = \Big\lceil{\pi^2\over 4(1-s)\,m^2}\Big\rceil.
\end{equation}
The use of the ``ceiling'' operators $\lceil\cdot\rceil$ in \eqref{sinc_constants}
ensure $N^-$ $N^+$ are integer valued, as required.

\begin{remark}
{\rm 
In using \eqref{cjw5}, we 
avoid the costly (and in many cases, inaccurate) precalculation of the eigenspectrum for the 
Laplacian over an arbitrary spatial domain $\Omega$ with Dirichlet condition $u\vert_\Gamma = g$. 
Even in cases where calculation of the eigenspectrum bears an acceptable computational 
cost, there still remains the outstanding question of just how much of the spectrum 
is required for computing the Laplacian by this method to acceptable accuracy.  For these
reasons, our equation \eqref{cjw5} is far more practical. 
}
\end{remark}

Rewriting \eqref{cjw2} as $v - \left (-\Delta \right )^{-s}\,k^2 (v+w)
= \left (-\Delta \right )^{-s}f$, we may write $v$ as a Kato--style
expansion
\begin{equation}
\label{cjw6}
v = {\sin s\pi \over \pi} m 
\sum_{\ell=-N^-}^{N^+} e^{(1-s)y_\ell} v_\ell
\end{equation}
and equate each of the $\ell$ terms to arrive at the coupled equation
\begin{equation}
\label{cjw7}
v_\ell - \left ( e^{y_\ell}-\Delta\right )^{-1}\,k^2\left (v + w\right ) = 
\left ( e^{y_\ell} - \Delta \right )^{-1}f,
\end{equation}
or
\begin{equation}
\label{cjw8}
\left ( e^{y_\ell} - \Delta \right ) v_\ell - 
k^2\left (v + w\right ) =  f.
\end{equation}
Observe that there are $N^- + N^+ + 1$ of these equations and that the 
$\ell$th equation fully couples the function $v_\ell$ into $w$ 
and all remaining functions $v_{\ell^\prime\ne\ell}$ of the expansion 
\eqref{cjw6}. Enforcement of the modified boundary condition
$v_\ell = 0$ guarantees enforcement of $v = 0$ via \eqref{cjw6}.  Hence, with the inclusion of \eqref{cjw4}, we have 
the complete differential problem statement consisting of 
a coupled system of $N^- + N^+ + 2$ equations with 
unknown functions $ v_{N^-},\ldots,v_{N^+}, w$ over the domain $\Omega$.

\section{Numerical Implementation}
\label{numerical}
The method of solution for equations \eqref{cjw4} and \eqref{cjw8}
(including corresponding boundary conditions) is to first transform 
the differential problem statement into an equivalent variational problem
statement for the appropriate infinite dimensional function spaces and 
then approximate its solution by the optimal one, in a Sobolev norm sense, 
taken from finite dimensional space of linear, nodal finite 
elements over some discretization.  In doing so, we introduce 
the test function $\xi_\ell$ and construct the weak form of \eqref{cjw8} for all
$\ell$ from $-N^-$ to $N^+$:
\begin{equation}
\label{cjw9}
\int_\Omega \left (-\xi_\ell\, \Delta v_\ell + e^{y_\ell}\,\xi_\ell v_\ell - k^2\, \xi_\ell \left ( v + w\right ) \right )\,{\text d}\Omega
= \int_\Omega \xi_\ell f\,{\text d}\Omega,
\end{equation}
recalling that $v$ is given by the expansion \eqref{cjw6}.  The test function $\zeta$ is 
used in the weak form 
of the Laplace equation \eqref{cjw4} as,
\begin{equation}
\label{cjw10}
\int_\Omega \nabla\zeta \cdot \nabla w\,{\text d}\Omega  = 0.
\end{equation}
Combining the left hand sides of \eqref{cjw9} and \eqref{cjw10}, the bilinear form $A(\cdot,\cdot)$ is 
therefore given as 
\begin{equation}
\label{cjw11}
\begin{aligned}
A(\left \{ \xi_\ell\right \},\zeta;\left \{v_\ell\right \},w)  = 
\sum_{\ell=-N^-}^{N^+} 
\int_\Omega & \left (\nabla\xi_\ell\cdot  \nabla v_\ell + e^{y_\ell}\,\xi_\ell v_\ell - k^2\, \xi_\ell \left ( v + w \right ) \right )
\,{\text d}\Omega \\
 & +
\int_\Omega \nabla\zeta \cdot \nabla w\,{\text d}\Omega,
\end{aligned}
\end{equation}
the combined right hand sides are denoted as
\begin{equation}
\label{cjw12}
F\left ( \left \{ \xi_\ell\right \},\zeta\right ) = 
\sum_{\ell=-N^-}^{N^+} 
\int_\Omega \xi_\ell f\,{\text d}\Omega,
\end{equation}
and $v$ is understood to be expanded in terms of $v_\ell$ according to \eqref{cjw6}.
The variational problem statement equivalent to the differential problem statement in \eqref{cjw4} and \eqref{cjw8} is 
therefore: Find $\left \{ v_\ell\right \}\in V_0; w \in V$, 
such that 
\begin{equation}
\label{cjw13}
A(\left \{ \xi_\ell\right \},\zeta;\left \{v_\ell\right \},w) = 
F\left ( \left \{ \xi_\ell\right \},\zeta\right )\quad\forall 
\left \{ \xi_\ell\right \},\zeta\in V_0,
\end{equation}
where $V$ is the space of $L^2$
functions on $\Omega$ 
with first order weak derivatives also in $L^2$ of $\Omega$ and inhomogeneous Dirichlet boundary conditions \eqref{cjw3}, and $V_0\subset V$ taking
homogeneous boundary conditions as in \eqref{cjw2}. 
The next step is to choose a finite--dimensional space $V_h \subset V$ from which the 
approximate solutions $v_h \approx v$ and $w_h\approx w$ will be drawn. 
To further simplify notation  we will drop the $h$ subscript and only re--introduce it as needed in relation to 
the true (weak) solutions $v$ and $w$.  

Let $\phi_1(\bm{x}), \phi_2(\bm{x}), \ldots, \phi_N(\bm{x})
$ be the basis functions in
$V_h$ as a function of spatial coordinate $\bm
x$, which in our implementation will be linear, nodal finite elements
.  We write in bold the column vector
$\bm{\phi}$ of basis functions $(\phi_1, \phi_2, \ldots,
\phi_N)^T$, and $\bm{v}_\ell=\left ( v_\ell^1,
  v_\ell^2,\ldots,v_\ell^N \right
)^T$.  By construction, it follows that $v_\ell(\bm{x}) =
\bm{\phi}^T\bm{v}_\ell = \sum_{i=1}^N \phi_i (\bm{x})
v_\ell^i$.  Likewise, $\bm{\xi}_\ell = \left (\xi_\ell^1,
  \xi_\ell^2,\ldots,\xi_\ell^N \right )^T$ and $\bm{w}=\left ( w_1,
  w_2,\ldots,w_N \right )^T$ yield $\xi_\ell =
\bm{\xi}_\ell^T\bm{\phi}$ and $w =
\bm{\phi}^T\bm{w}$, respectively.  As such, construction of the linear
system \eqref{cjw13} requires the following block matrices built by
volume integration of the basis functions and their spatial
derivatives:
\begin{equation}
\label{cjw14}
{\bold K}    = \int_\Omega \left ( \nabla\bm{\phi} \right )^T \left ( \nabla \bm{\phi}\right ) \,{\text d}\Omega, 
\quad {\bold M}_1  = \int_\Omega \bm{\phi}\bm{\phi}^T\,{\text d}\Omega , 
\quad \mbox{and}\quad {\bold M}_2  = -\int_\Omega k^2\, \bm{\phi}\bm{\phi}^T\,{\text d}\Omega 
\end{equation}
where the integrands are understood to be outer products, each yielding a symmetric matrix of dimension $N\times N$.  To
simplify notation, introduce the coefficients $c_\ell = e^{y_{\ell-N^-}}$ and $d_\ell = {1\over \pi} (\sin s\pi)\, m \, 
e^{(1-s)y_{\ell-N^-}}$, matrix ${\bold A}_\ell = {\bold K} + c_\ell{\bold M}_1 + d_\ell{\bold M}_2$, and sum $L = N^- + N^+$ 
so that we may compactly write $A(\left \{ \xi_\ell\right \},\zeta; \left \{v_\ell\right \},w)$ as
\begin{equation}
\label{cjw15}
\left ( \bm{\xi}^T_{-N^-} \cdots \bm{\xi}^T_{N^+} \,\bm{\zeta}^T \right)
\left (
\begin{array}{cccccc}
{\bold A}_0      & d_1\,{\bold M}_2 & d_2\,{\bold M}_2 & \cdots & d_L{\bold M}_2 & {\bold M}_2 \\
d_0\,{\bold M}_2 & {\bold A}_1      & d_2\,{\bold M}_2 & \cdots & d_L{\bold M}_2 & {\bold M}_2 \\
d_0\,{\bold M}_2 & d_1\,{\bold M}_2 & {\bold A}_2      & \cdots & d_L{\bold M}_2 & {\bold M}_2 \\
\vdots           & \vdots & \vdots & \ddots & \vdots & \vdots \\ 
d_0\,{\bold M}_2 & d_1\,{\bold M}_2 & d_2\,{\bold M}_2 & \cdots & {\bold A}_L    & {\bold M}_2 \\
{\bold 0} & {\bold 0} & {\bold 0} & \cdots & {\bold 0}    & {\bold K} \\
\end{array}
\right )
\left (
\begin{array}{c}
\bm{v}_{-N^-}\\
\bm{v}_{1-N^-}\\
\bm{v}_{2-N^-}\\
\vdots\\
\bm{v}_{N^+}\\
\bm{w}\\
\end{array}
\right ).
\end{equation}
Lastly, the right hand side of \eqref{cjw13} follows as
\begin{equation}
\label{cjw17}
\left ( \bm{\xi}^T_{-N^-} \cdots \bm{\xi}^T_{N^+} \,\bm{\zeta}^T \right)
\left (
\begin{array}{c}
\bm{f}\\
\bm{f}\\
\bm{f}\\
\vdots\\
\bm{f}\\
\bm{0} \\
\end{array}
\right )
\end{equation}
with  column vector  $\bm{f} = \int_\Omega \bm{\phi} f\,{\text d}\Omega$.
In equating \eqref{cjw15} with \eqref{cjw17}
as required by the variational problem statement \eqref{cjw13}, we see that the coefficient vector 
$\left ( \bm{\xi}^T_{-N^-} \cdots \bm{\xi}^T_{N^+} \,\bm{\zeta}^T \right)$ is common to both the 
left and right hand sides, and may therefore be divided out, thus leaving a $N(L+2)\times N(L+2)$ system 
of linear equations for the unknown coefficients $\left ( \bm{v}^T_{-N^-} \cdots \bm{v}^T_{N^+} \,\bm{w}^T \right)$
which holds for all functions $\left \{ \xi_\ell \right \}, \zeta \in V_h$.  Upon solution of the linear system, aggregation
of the coefficient vectors $\bm{v}_\ell$ according to \eqref{cjw6} plus the vector $\bm{w}$ completes the sum $v + w$, 
which we recognize as the discrete, approximate solution to the original differential equation \eqref{cjw1}.

Because the matrix in \eqref{cjw15} is complex--valued, large and nonsymmetric, the solution strategy for the linear system equating
\eqref{cjw15} and \eqref{cjw17} must be carefully chosen for scalability and economy of compute resources. As such, we solve this linear
system using stabilized bi-conjugate gradients (BiCG-STAB) \cite{van_der_Vorst:1992}: The algorithm is easily parallelizable; has a minimum number 
of working vectors; and requires only two matrix--vector products per iterative step. The latter is especially important for reducing 
computational resource burdens because these products can be computed cheaply and quickly ``on the fly'' as needed and without 
explicit storage of entire system matrix.  Notice, however, that the matrix in \eqref{cjw11} is block dense and a large number of floating 
point operations is required for a single matrix--vector multiply -- operations which may significantly increase the time required to 
perform the multiplications.  To remedy this, we modify the variational formulation \eqref{cjw13} to include the function $v$ in addition
to the vectors $v_\ell$ and $w$, and augment $A(\cdot)$ by weak enforcement of the compatibility expansion \eqref{cjw6} between
vectors $\bm{v}_\ell$ and $\bm{v}$.   That is, in addition to $\{v_\ell\}$ and $w$, we introduce the 
additional unknown $v$ and find  
$\left \{ v_\ell\right \},w$ and  $v$  such that 
\begin{equation}
\label{cjw18}
{\tilde A}(\left \{ \xi_\ell\right \},\zeta,\eta;\left \{v_\ell\right \},w,v) = 
{\tilde F}\left ( \left \{ \xi_\ell\right \},\zeta,\eta \right )\quad\forall 
\left \{ \xi_\ell\right \},\zeta, \text{ and } \eta,
\end{equation}
where
\begin{equation}
\label{cjw19}
\begin{aligned}
{\tilde A}(\left \{ \xi_\ell\right \} ,\zeta,\eta;& \left \{v_\ell\right \},w,v)  =  \\
& \sum_{\ell=N^-}^{N^+} 
 \int_\Omega  \left (\nabla\xi_\ell\cdot \nabla v_\ell + e^{y_\ell}\,\xi_\ell v_\ell - k^2\, \xi_\ell \left ( v + w \right ) 
+ \nabla\zeta\cdot\nabla w
\right )
\,{\text d}\Omega \\
&\quad\quad\quad - {\sin s\pi \over \pi} m \sum_{\ell=-N^-}^{N^+} e^{(1-s)y_\ell} \int_\Omega \eta\,v_\ell\,{\text d}\Omega \quad +\quad
\int_\Omega \eta\,v\,{\text d}\Omega
\end{aligned}
\end{equation}
and
\begin{equation}
\label{cjw20}
{\tilde F}\left ( \left \{ \xi_\ell\right \},\zeta,\eta \right ) = 
\sum_{\ell=N^-}^{N^+} 
\int_\Omega \xi_\ell f\,{\text d}\Omega.
\end{equation}
The resulting sparse linear system is thus,
\begin{equation}
\label{cjw21}
\left (
\begin{array}{ccccccc}
\tilde{\bold A}_0    & {\bold 0}   & \cdots    & \cdots  & {\bold 0} & {\bold M}_2 & {\bold M}_2 \\
{\bold 0}      & \tilde{\bold A}_1 & {\bold 0} & \cdots  & {\bold 0} & {\bold M}_2 & {\bold M}_2 \\
 \vdots      &  & \ddots &  & \vdots & {\bold M}_2 & {\bold M}_2 \\
 {\bold 0}      &  \cdots &  {\bold 0}  & \tilde{\bold A}_{L-1} & {\bold 0} & {\bold M}_2 & {\bold M}_2 \\
 {\bold 0}      &  \cdots &  \cdots     & {\bold 0}  & \tilde{\bold A}_L  & {\bold M}_2 & {\bold M}_2 \\
 {\bold 0}      &  \cdots &  \cdots     & \cdots  & {\bold 0}  & {\bold K} & {\bold 0} \\
 -d_0{\bold M}_1      &  \cdots &  \cdots  & \cdots   & -d_L{\bold M}_1  & {\bold 0}  &  {\bold M}_1 \\
\end{array}
\right )
\left (
\begin{array}{c}
\bm{v}_{-N^-}\\
\bm{v}_{1-N^-}\\
\bm{v}_{2-N^-}\\
\vdots\\
\bm{v}_{N^+}\\
\bm{w}\\
\bm{v}\\
\end{array}
\right )
= 
\left (
\begin{array}{c}
\bm{f}\\
\bm{f}\\
\bm{f}\\
\vdots\\
\bm{f}\\
\bm{0}\\
\bm{0}\\
\end{array}
\right )
\end{equation}
with 
$\tilde{\bold A}_\ell = {\bold K} + c_\ell{\bold M}_1$.
We refer to the sparsified system of linear equations in \eqref{cjw21} as the {\it v-formulation} for the discretized, variational 
form of original differential equation \eqref{cjw1}. 
Inspection of the prefactors $d_\ell$ from the Kato expansion, however, suggests 
that their exponential decay with respect to $\ell$ may lead to ill-conditioning of the coefficient matrix 
in \eqref{cjw21} by their presence in the last block--row.  As such, we recast \eqref{cjw21} as the {\it scaled v-formulation}:
\begin{equation}
\label{cjw22}
\left (
\begin{array}{ccccccc}
\tilde{\bold A}^\prime_0    & {\bold 0}   & \cdots    & \cdots  & {\bold 0} & {\bold M}_2 & {\bold M}_2 \\
{\bold 0}      & \tilde{\bold A}^\prime_1 & {\bold 0} & \cdots  & {\bold 0} & {\bold M}_2 & {\bold M}_2 \\
 \vdots      &  & \ddots &  & \vdots & {\bold M}_2 & {\bold M}_2 \\
 {\bold 0}      &  \cdots &  {\bold 0}  & \tilde{\bold A}^\prime_{L-1} & {\bold 0} & {\bold M}_2 & {\bold M}_2 \\
 {\bold 0}      &  \cdots &  \cdots     & {\bold 0}  & \tilde{\bold A}^\prime_L  & {\bold M}_2 & {\bold M}_2 \\
 {\bold 0}      &  \cdots &  \cdots     & \cdots  & {\bold 0}  & {\bold K} & {\bold 0} \\
 -{\bold M}_1      &  \cdots &  \cdots  & \cdots   & -{\bold M}_1  & {\bold 0}  &  {\bold M}_1 \\
\end{array}
\right )
\left (
\begin{array}{c}
\bm{v}^\prime_{-N^-}\\
\bm{v}^\prime_{1-N^-}\\
\bm{v}^\prime_{2-N^-}\\
\vdots\\
\bm{v}^\prime_{N^+}\\
\bm{w}\\
\bm{v}\\
\end{array}
\right )
= 
\left (
\begin{array}{c}
\bm{f}\\
\bm{f}\\
\bm{f}\\
\vdots\\
\bm{f}\\
\bm{0}\\
\bm{0}\\
\end{array}
\right )
\end{equation}
with 
$\tilde{\bold A}^\prime_\ell = \left ( {\bold K} + c_\ell{\bold M}_1\right )/d_\ell$ and $\bm{v}^\prime_\ell =
 \bm{v}_\ell d_{\ell + N^-}$.
In the scaled system \eqref{cjw22}, the Kato scale factors $d_\ell$ are implicit in the unknown vectors 
$\bm{v}^\prime_\ell$ and act within
block diagonal matrices $\tilde{\bm A}^\prime_\ell$ alone.  

As a closing remark on the theory and algorithms just described,
observe that in \eqref{cjw4}, \eqref{cjw21} and \eqref{cjw22}, the
solution of the scalar Laplacian equation for $w$ is fully decoupled
from the solution for $v$ and $v_\ell$.  Hence, one option for solving
the full system of equations is a two--step procedure, where first the
solution for $w$ is obtained, and then used as a sourcing term for the
remaining $v_\ell$ equations.  We have, instead, chosen to solve the
full system simultaneously.  This has some advantages.  First, in
looking ahead to the implementation of Robin boundary conditions
(e.g. a Sommerfeld radiation condition), we anticipate that the
Laplacian equations will couple directly into the $v$ (or, equivalently,
$v_\ell$) equations, which would consequently eliminate the
convenience of solving for $w$ a priori.  We wish this coupling to
modify our existing algorithm/code structure as little as possible and
therefore retain the Laplacian equation for $w$ in the full system
matrix.  Second, including the Laplacian comes at an increased cost of
only $N$ degrees of freedom on top of the existing cost of $L\,N$ for
the $v_\ell$ equations.  Because $L$ is typically on the order of 100
or more for adequately refined meshes (\figref{nquad}), this added
cost is objectively minimal.  Lastly, looking further ahead toward PDE
constrained optimization where we might invert for $s$ or
$\sigma_{\alpha,\zeta}$, or for the design problem (e.g. optimal
sensor sensor placement), it is more convenient to create and
solve for the corresponding adjoint objects.

\begin{figure}
\centering \includegraphics[width=0.5\textwidth]{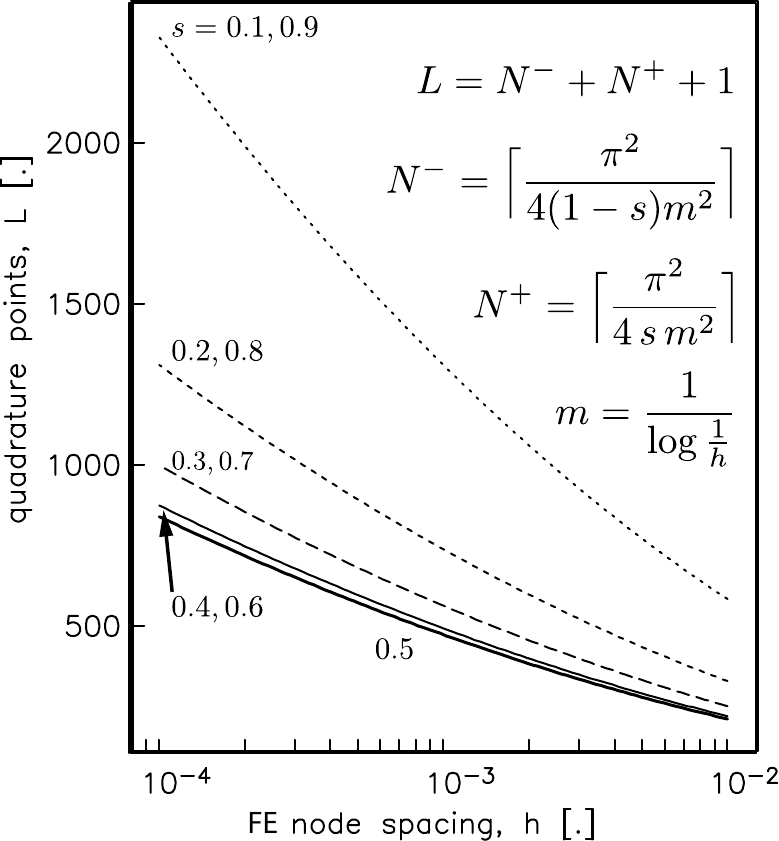}
\caption{Total number of quadrature points $L$ as a function of Laplacian exponent $s$ and 
node spacing $h$. Sinc quadrature summation is over the range of indices $\ell = -N^-,\ldots, N^+$.}
\label{nquad}
\end{figure}

\subsection{Non--locality of the fractional Laplacian operator}

We next provide insight as to why equations \eqref{cjw5},
\eqref{spectral} are nonlocal operators. As pointed out in
\cite{Song_and_Vondracek:2008,Caffarelli_and_Stinga:2016,Antil_etal:2017,HAntil_MWarma_2019a},
the spectral fractional Laplacian with zero boundary conditions can be
equivalently written as:
  \begin{equation}\label{eq:equivflap0}
    (-\Delta_0)^s u(\bm{x}) = \int_\Omega ( u(\bm{x}) - u(\bm{x}^\prime) ) 
\mathcal{K}_s(\bm{x},\bm{x}^\prime)\,\text{d}\Omega^\prime
+ u(\bm{x}) B_s(\bm{x})
  \end{equation}
where $\mathcal{K}_s$ and $B_s$ are appropriate Kernel functions. 
 One can see from the equivalent definition that, in order to evaluate $(-\Delta_0)^s u$
at a point $\bm{x} \in \Omega$, we need information about $u$ on the entire domain $\Omega$,
thus making $(-\Delta_0)^s$ a nonlocal operator. Moreover, let $\mathcal{O}$ be an
open set contained in $\Omega$ such that $u \equiv 0$ on $\mathcal{O}$. Then a classical
Laplacian implies $-\Delta u(\bm{x}) = 0$ for all $\bm{x} \in \Omega$. 
However, this is not the case 
when we deal with $(-\Delta_0)^s$. Indeed, let $\bm{x} \in \mathcal{O}$ and let use the equivalent
definition \eqref{eq:equivflap0}, since $u \equiv 0$ on $\mathcal{O}$, we obtain that 
  \begin{align}
    (-\Delta_0)^s u(\bm{x}) &=  \int_\Omega \left ( 
u(\bm{x}) - u(\bm{x}^\prime) \right ) \mathcal{K}_s(\bm{x},\bm{x}^\prime) {\text d}\Omega^\prime 
\nonumber \\
               &= \int_{\mathcal{O}} \left ( u(\bm{x}) - u(\bm{x}^\prime) \right ) 
\mathcal{K}_s(\bm{x},\bm{x}^\prime)\, {\text d}\Omega^\prime + 
              \int_{\Omega \setminus \mathcal{O}} \left ( - u(\bm{x}^\prime) \right ) 
\mathcal{K}_s(\bm{x},\bm{x}^\prime) \,{\text d}\Omega^\prime  \nonumber \\
               &=  \int_{\Omega \setminus \mathcal{O}} 
\left ( - u(\bm{x}^\prime) \right ) \mathcal{K}_s(\bm{x},\bm{x}^\prime) \,{\text d}\Omega^\prime 
  \end{align}  
which is not necessarily zero. This is unlike the local case.

\section{Numerical Verification}

To verify the implementation of  the fractional Helmholtz equations with inhomogeneous
Dirichlet boundary conditions we adopt the Method of Manufactured 
Solutions (MMS) \cite{Roache:1998, Salar_and_Knupp:2000}. In the MMS method, a proposed solution 
is substituted into the governing differential equation, after which the corresponding
boundary conditions and sourcing functions are derived.  Upon discretization, the
recovered numerical solution is then compared to the known analytical solution.  The 
MMS solution used here over the interval $x\in[0,1]$ takes the following form:
$v = \sin(2\pi x), w = 1 \mapsto u=1+\sin(2\pi x)$.  Inspection of
\eqref{cjw4} results in setting $g(0)=g(1) = 1$, whereas recognizing 
that $\sin(2\pi x)$ is an eigenfunction for homogeneous $(-\Delta)^s$ 
results in $f= \left ((2\pi)^s - k^2\right )\sin(2\pi x) - k^2$ according to 
\eqref{cjw2}.  Hence, we have constructed for arbitrary $s$ the requisite source terms and boundary 
conditions for our posited solution $u$ solving an inhomogeneous
fractional Helmholtz with non-zero Dirichlet boundary conditions.  Note that MMS solution $u$ is 
independent of both $s$ and $k$ and is thus powerful test of fractional Helmholtz algorithm.

Numerical evaluation of the MMS problem just described is done using linear, nodal 
finite elements with uniform node spacing for the case of $s = 0.25$ and $k$ arbitrarily 
set to unity. Linear system \eqref{cjw22} is solved to high accuracy using the stabilized bi-conjugate gradient
\cite{van_der_Vorst:1992} iterative scheme with simple Jacobi scaling to a tolerance of $10^{-16}$ reduction 
in normalized residual.  Over the range of node spacing $0.001 \le h \le 0.01$ the MMS solution 
shows the expected $h^2$ convergence in error between the recovered finite element and known 
analytic solutions (\figref{mms_s025}) in $L^2(\Omega)$-norm.  For reference, the size $N_{total}$ of the linear system 
\eqref{cjw22} grows roughly as $N^{1.4}$ over the corresponding range in $h$, resulting, for example, in $L=629$ quadrature points for 
$N=1001$ finite element nodes and a total of $N_{total} =631631$ unknowns in the linear system \eqref{cjw22}.  Convergence of 
the bi--conjugate gradient residual error as a function of iteration count (\figref{bicg}) is generally well 
behaved, with only minor localized excursions from monotonicity.  Furthermore, the error in simultaneously solving each 
of the three sets of coupled equations -- fractional Helmholtz for $v_\ell$; Laplacian for $w$; and, compatibility between
$v$ and $v_\ell$ -- decreases synchronously with iteration count, with error for the compatibility equation 
approximately a factor 100 less than the error for the remaining two.  

\begin{figure}
\centering \includegraphics[width=0.6\textwidth]{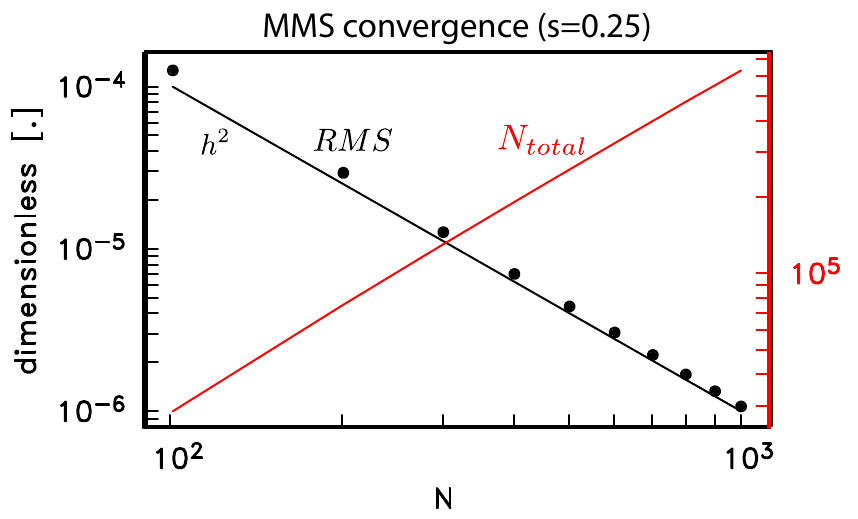}
\caption{Convergence of MMS solution $u=\sin(2\pi x) + 1$ for $s=0.25$ as a function of mesh size $N$ 
with corresponding node spacing $(N-1)^{-1}$.  In symbols is the RMS residual; black lines, the
curve $h^{2}$; and in red, the total number of degrees of freedom in the discretized linear system \eqref{cjw22}.
Linear system \eqref{cjw22} is solved using BiCG-STAB to a tolerance of $10^{-16}$ in RHS-normalized residual.  }
\label{mms_s025}
\end{figure}

\begin{figure}
\centering \includegraphics[width=0.5\textwidth]{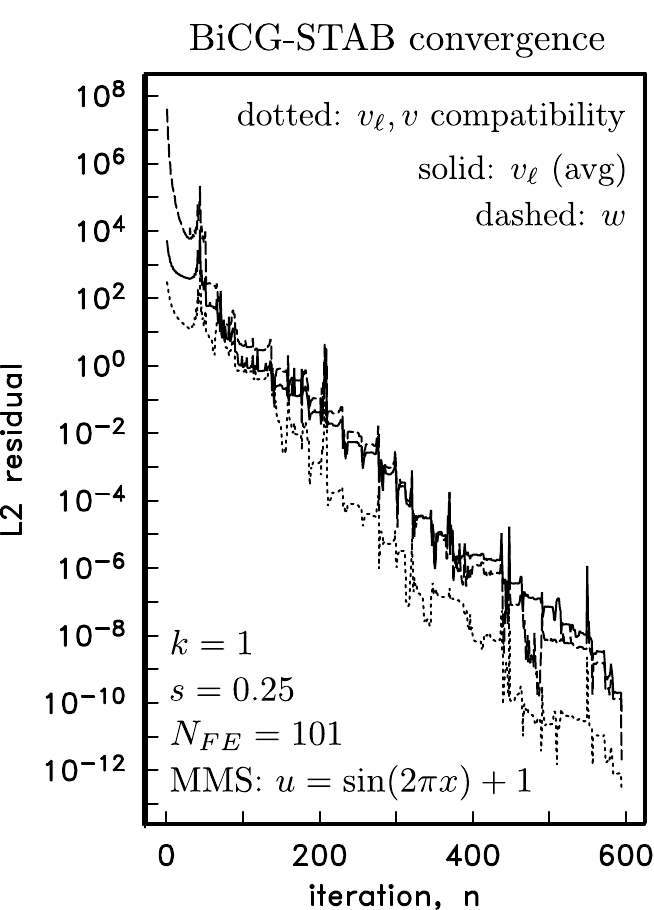}
\caption{Convergence of BiCG-STAB algorithm with Jacobi preconditioning for the $N=101$ MMS solution in Figure 1.  
In the solid line is the average residual for equations in blocks $0,\ldots,L$ in \eqref{cjw22} corresponding
to fractional Helmholtz equations on $v_\ell$;  
dashed, the residual for equations in block $L+1$ corresponding to solution of the Laplace equation for $w$; and
dotted, the residual for the final $(L+2)$ block of equations enforcing compatibility between $v$ and $v_\ell$. }
\label{bicg}
\end{figure}

Lastly, we confirm that the choice
$m=1/\log{1\over h}$ for quadrature spacing (and by extension, the number $L$ of $v_\ell$ equations) is nearly 
optimal by examining the effect on RMS of varying $m$.  As a representative example, the $N=101, s=0.25$ 
discretization of the MMS problem is solved for a range of $m$ values around $1/\log{1\over h}$. In this example the 
asymptotic limit for minimum RMS value is achieved at $m$ roughly 90\% of its optimal value, where the 
asymptotic limit is driven by the error of the finite element discretization itself (\figref{sinc}).  In contrast, 
choices of $m$ larger than the optimal value result in a rapidly increasing RMS, consistent with the 
exponential convergence of quadrature error reported elsewhere \cite{ABonito_JEPasciak_2015a}.  

\begin{figure}
\centering \includegraphics[width=0.75\textwidth]{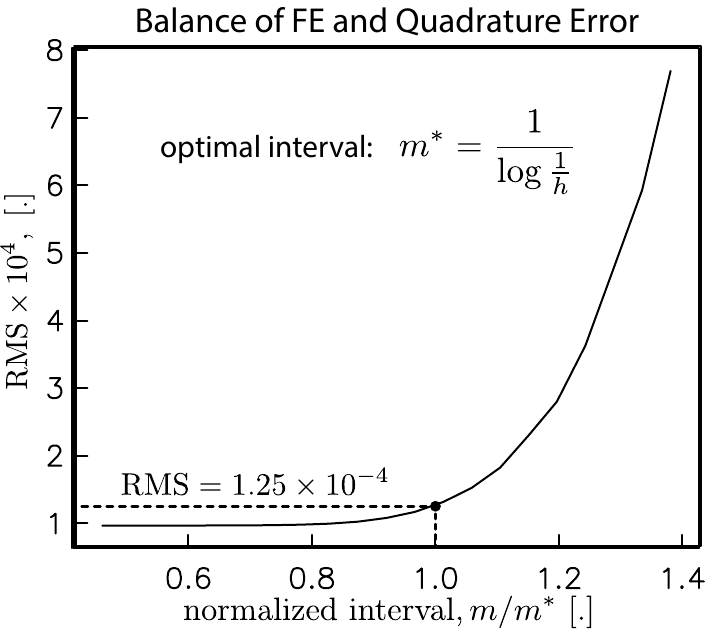}
\caption{Convergence of RMS error as a function of quadrature spacing $m$ for the MMS problem with $N=101, h=0.01$ 
finite element nodes (see \figref{mms_s025}). Optimal quadrature interval $m^*$ is given by \eqref{sinc_constants} yields $\text{RMS}=1.25\times 10^{-4}$, 
a value close to the asymptotic limit when $m < m^*$.  Note the rapidly increasing RMS error as $m^* < m$.}
\label{sinc}
\end{figure}


\section{Numerical Results}
\par\bigskip

Our fractional Helmholtz system is numerically demonstrated in the
context of magnetotellurics (MT).  This is a geophysical surveying
method that measures naturally occurring, time-varying magnetic and
electric fields.  Resistivity estimates of the subsurface can be
derived, from the very near surface to hundreds of  kilometers, that
are applied to subsurface characterization for geoscience application,
such as hydrocarbon extraction, geothermal energy harvesting, and
carbon sequestration, as well as studies into Earth's deep tectonic history.  
The MT signal is caused by the interaction of
the solar wind with the earth’s magnetic field (lower frequencies less
than 1 Hz) and world-wide thunderstorms, usually near the equator
(higher frequencies greater than 1 Hz).  Figure (\ref{sketch})
provides a conceptual diagram of MT in which the ionosphere is the 
electromagnetic source (see Section 2) for inducing currents 
in the subsurface.  Because the magnitude of this source current
is unknown, the fundamental quantity for MT analysis is the 
impedance tensor mapping electric correlated electric and magentic fields.
Computed in the frequency domain, the impedance tensor is an estimate
of the Earth ``filter'' mapping magnetic to electric fields -- in other
words, it is an expression of Earth's conductivity distribution. 
Common in preliminary MT analysis is the assumption of locally 1D (depth
dependent) electrical structure and excitation by a vertically 
incident plane wave, such as described in Section 2.  We adopt these modest
assumptions in our investigation of MT data -- in particular, data
collected by the decadal, trans--continental USArray/Earthscope 
project \cite{Earthscope} -- and find examples where MT data 
is consistent predicted impedances for a fractionally--diffusing 
electromagnetic Earth.

Because of the novelty in applying fractional derivative concepts to
electromagnetic geophysics, the first question that draws our
attention is simply: How does a fractionally diffusing field, as
described by \eqref{fhelm} and \eqref{fhelm_kappa}, compare to a field
derived from the classical Helmholtz equation?  To address this
question we solve \eqref{fhelm_kappa} on the dimensionless unit
interval $\zeta\in[0,1]$ with unit amplitude Dirichlet conditions
$u(0)=\sqrt{2\text{i}}$ and $u(1)=0$ on the horizontal electric field
and choose the dimensioned scaling factor $z^*=1000$ m to represent
the physical domain $z\in[0,1000]$ m.  Choice of homogeneous Dirichlet
condition at $\zeta = 1$ is commonly known as ``perfectly conducting''
boundary condition, representing the presence of an infinitely
conductive region for $\zeta > 1$, but is used here strictly out of
computational convenience. Scattering from this interface back to
Earth's surface $\zeta=0$ will be negligible as long as the frequency
$\omega$ in Equation \eqref{fhelm_kappa} is sufficiently high that the
electric field at depth is essentially zero.  The unit interval is
discretized with 501 evenly distributed nodes, on which the electric
field is drawn from the finite dimensional vector space of linear
nodal finite elements. Hence, node spacing is $h=0.002$, which, when
$s=0.7$ for example, leads to $N^-=318$ and $N^+=137$ according to
\eqref{sinc_constants} and a linear system \eqref{cjw22} with
$501(3 + N^- + N^+) = 229458$ equations.  Comparable to the error
tolerances on the BiCG-STAB solver specified previously for the MMS
problem, the iterative sequence is terminated once the normalized
residual is reduced by $10^{-12}$ over its starting value.

\begin{figure} 
\centering \includegraphics[width=0.7\textwidth]{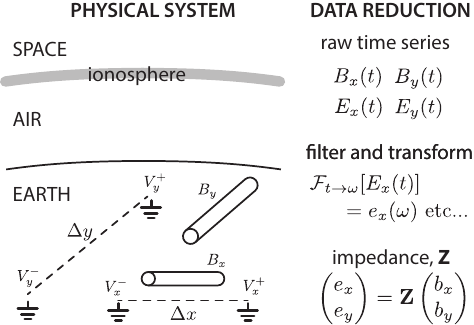}
\caption{ Overview of magnetotelluric experiment and data reduction. (left) Collocated
time series of horizontal electric field, measured by pairs of grounded electrodes, 
and magnetic field, measured by induction coils or fluxgates, measure Earth's inductive
response to ionspheric source currents.  (right) Time series are windowed, filtered and transformed
into the frequency domain, from which the impedance tensor is estimated, containing information 
on the distribution of electricical conductivity variations in Earth's subsurface \cite{Chave_and_Jones:2012}.
Because high--frequency fields decay more rapidly with depth than low frequency fields, frequency
can loosely be interpreted as a proxy for depth, and hence an impedance spectrum is a coarse
measure of the local, depth variations in electrical conductivity.}
\label{sketch}
\end{figure}
 
The horizontal electric fields in \figref{decay} show depth--dependent behavior 
that is clearly also $s$-dependent: increased curvature in the near--surface and decreased
curvature at depth in comparison with their classical $s=1.0$ counterpart. This 
suggests that the effect of the fractional Laplacian in \eqref{fhelm} and \eqref{fhelm_kappa} over a 
uniform $\sigma_{\alpha,\zeta}$ Earth model is, at first blush, in some ways similar to that of a 
classical Laplacian over a layered Earth which is conductive in the near surface
and resistive at depth. However, closer inspection of the fractional response 
(see, for example the $s=0.60$ curves) reveals that the damped oscillations, characteristic of classical Helmholtz,
are simply not present as $s$ decreases from unity, and instead then are replaced with a steady 
non--oscillatory decay with depth.

\begin{figure} 
\centering \includegraphics[width=0.5\textwidth]{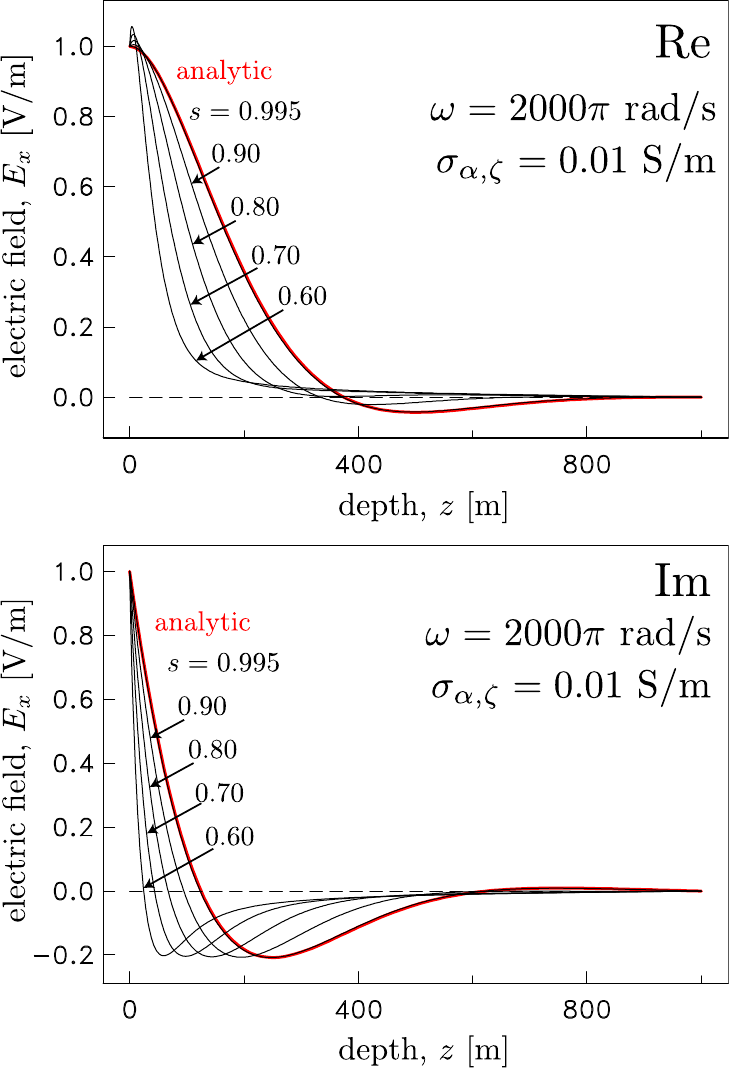}
\caption{ For a range of fractional exponent values $s=0.6, 0.7, 0.8, 0.9, 0.995$,
decay of unit--amplitude real (top) imaginary (bottom) components of horizontal 
electric field $E_x = u$ as a function of depth $z$ into a uniform 
$\sigma_{\alpha,\zeta} = 0.01$ S/m medium at frequency $f=1$ kHz, 
corresponding to dimensionless wavenumber 
$\kappa \approx 8.89 $. In red is the analytic solution for the corresponding classical $s=1$ Helmholtz. 
See text for additional details on boundary conditions and scaling 
to the physical domain from the dimensionless unit interval.  }
\label{decay}
\end{figure}

There is a dramatic manifestation of this fractional Helmholtz response in observable magnetotelluric data 
through calculation of the impedance spectrum (\figref{sounding}). Amplitude of
the impedance spectrum, reported here as the familiar apparent resistivity 
\begin{equation}
\rho_a = {1\over\omega\mu_0} \Big\vert {E_x \over H_y} \Big\vert^2_{z=0} = \omega \mu_0 (z^*)^2 
\Big\vert{u\over\partial_\zeta u}\Big\vert^2_{\zeta = 0}
\label{appresis}
\end{equation}
and complex phase angle $\theta$  of the  ratio $-u/\partial_\zeta u$, show
a clear $s$--dependence at frequencies above 1 Hz. Decay of the apparent resistivity as 
frequency approaches zero can be understood as a consequence of the perfect electric
conductor boundary condition at $z=z*$, where at these low frequencies the reciprocal
wavenumber ${1\over \kappa }>> z^*$ and hence the apparent resistivity approaches that of the 
perfect conductor, zero, in the region $z > z^*$.  Furthermore, in the limit of zero frequency,
the fractional Helmholtz equation asymptotes to the fractional Laplacian equation (analagous
to \eqref{cjw3}) with inhomogeneous Dirichlet boundary conditions, whose solution 
has already been established \cite{Antil_etal:2018a} as equivalent to the classical 
Laplacian equation, leaving the ratio ${-u/\partial_\zeta u} =1$, or equivalently $\theta=0$.

\begin{figure} 
\centering \includegraphics[width=0.5\textwidth]{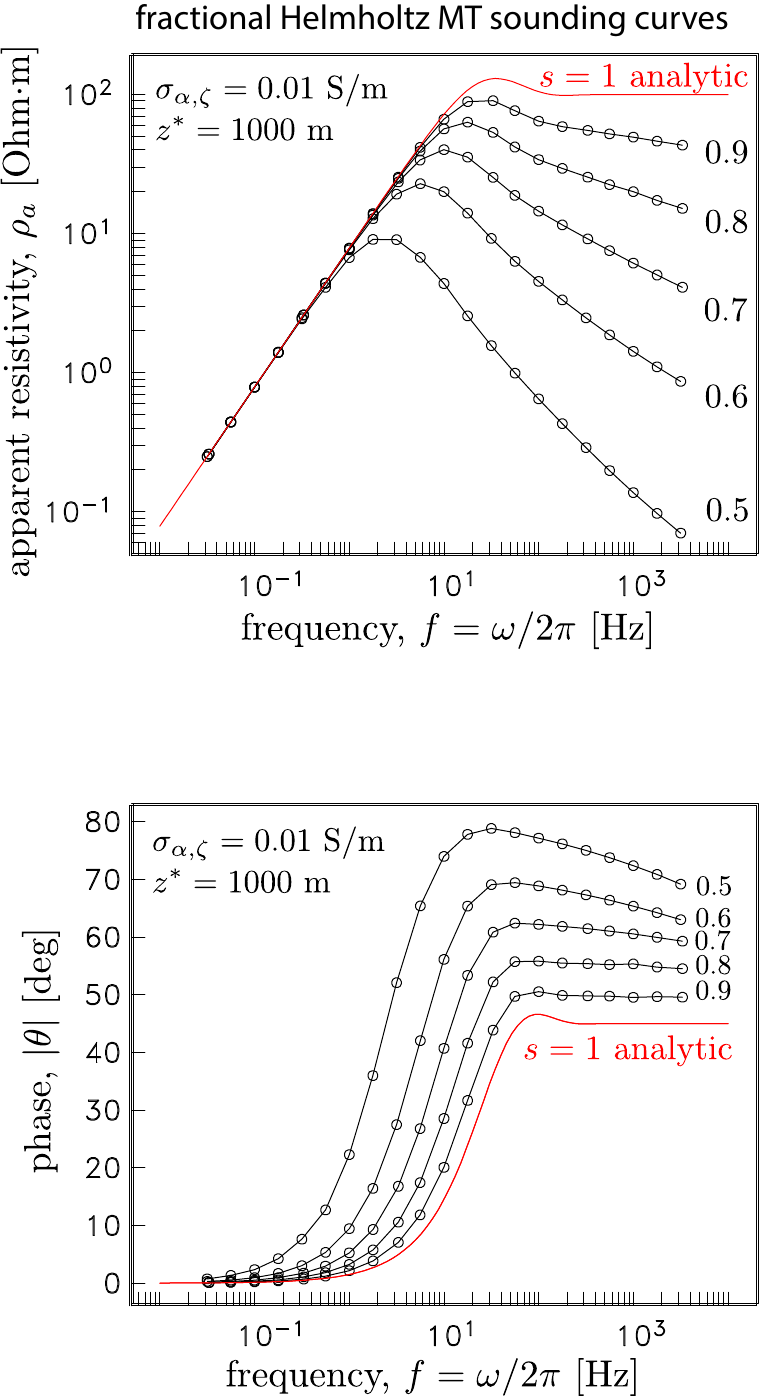}
\caption{Magnetotelluric sounding curves for a uniform $\sigma_{\alpha,\zeta}$ Earth model 
over the depth domain $z\in[0,z^*]$ for a range of fractional exponent values $s=0.5, \ldots, 1.0$ with
a perfect conductor boundary condition at $z=z^*$.
Apparent resistivity (top); complex phase (bottom).  See text for description. In red are
the classical $s=1$ Helmholtz solutions, computed analytically.  }
\label{sounding}
\end{figure}

The decrease in apparent resistivity at high frequencies when $s\ne 1$ can further
be understood by examination of the electric field gradient at $z=0$ (\figref{fvsc}). Although
there is a slight decrease in the vertical gradient of the imaginary component of 
electric field when $s\ne 1$, the magnitude of the real component increases dramatically 
in comparison to the $s=1$ case. This overall rise in vertical gradient at the 
air/earth interface for a fractional Earth model decreases the value of the quotient 
in \eqref{appresis}, thereby leading to a decreased estimate of the apparent resistivity at large frequencies.

\begin{figure} 
\centering \includegraphics[width=0.5\textwidth]{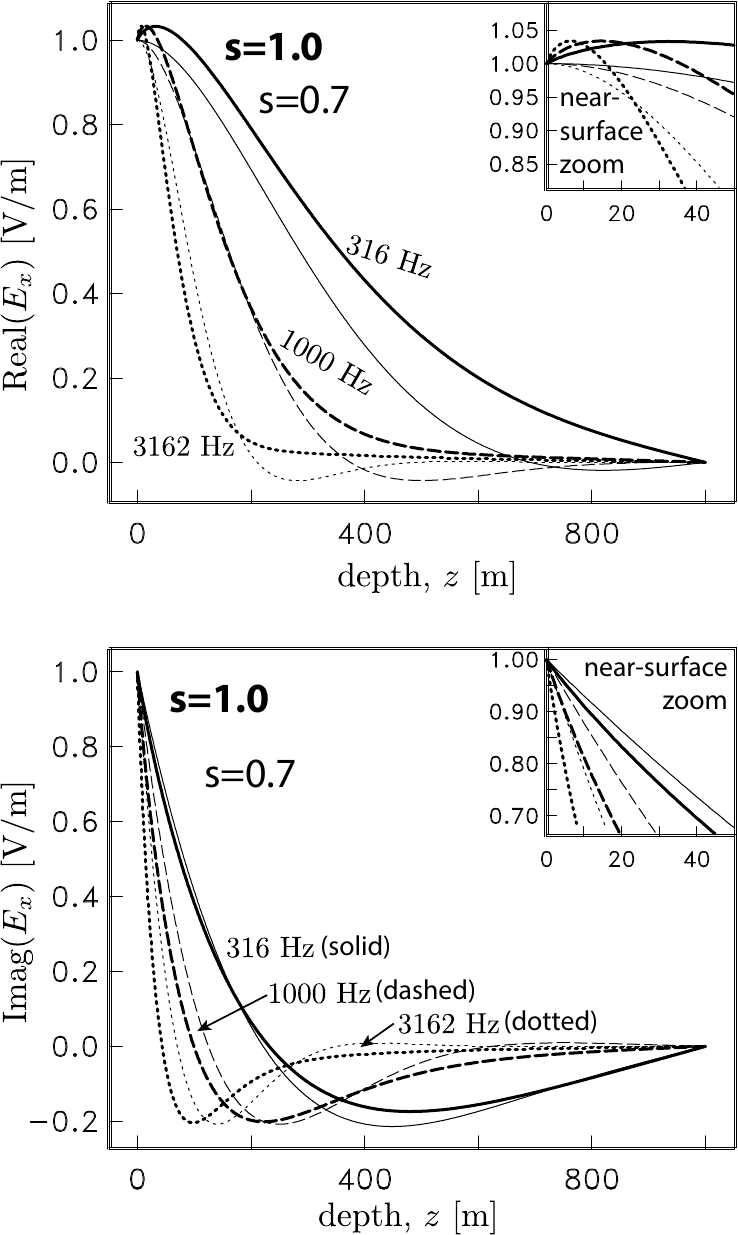}
\caption{Real (top) and imaginary (bottom) components of horizontal electric field as a function 
of depth in a uniform 0.01 S/m Earth underlain by a perfect conductor for frequencies $f=316, 1000$ and 3162 Hz, corresponding
to the high--frequency region of the magnetotelluric apparent resistivity spectrum (\figref{sounding}) with
approximate $s$ dependent power law behavior.   Curves for classical $s=1$ (heavy lines) and fractional $s=0.7$ response
(light lines) are shown.   The decrease in apparent resistivity is evidently due to the strongly increased vertical gradient of 
Real component of electric field at the air/Earth interface $z=0$ for fractional Helmholtz.  Recall that
from Eq (4.1) that the vertical gradient of electric fields resides in the denominator of the 
of the apparent resistivity estimator. }
\label{fvsc}
\end{figure}

\subsection{Validation through USArray data}

We have made progress towards validating our hypothesis of
``fractional Helmholtz leading to new geophysical interpretation''
though geophysical insight of numerical experiments.  Results suggests
complex material properties in the subsurface exposed to
electromagnetics energy exhibit conductive behavior near surface,
resisistive at depth. Additional evidence of superdiffusice behavior
can be observed from MT data at the USArray station NW Kansas
City. Apparent resistivity and phase angle data from USArray MT
station for KSP34 located NW of Kansas City, KS, USA show show similar
non-local behavior as our numerical experiments.  Figure (\ref{KSP})
shows apparent resistivity and phase angle versus frequency, as well as
resistivity versus depth, which exhibit non-local characteristics in
the subsurface geology. An in-depth study of the geology in the Kansas
City region would futher endorse our observations but is beyond the
scope of this paper.  These field data correlations however provide
further motivation to support additional algorithmic development for
fractional electromagnetics.

\begin{figure} 
\centering \includegraphics[width=1.0\textwidth]{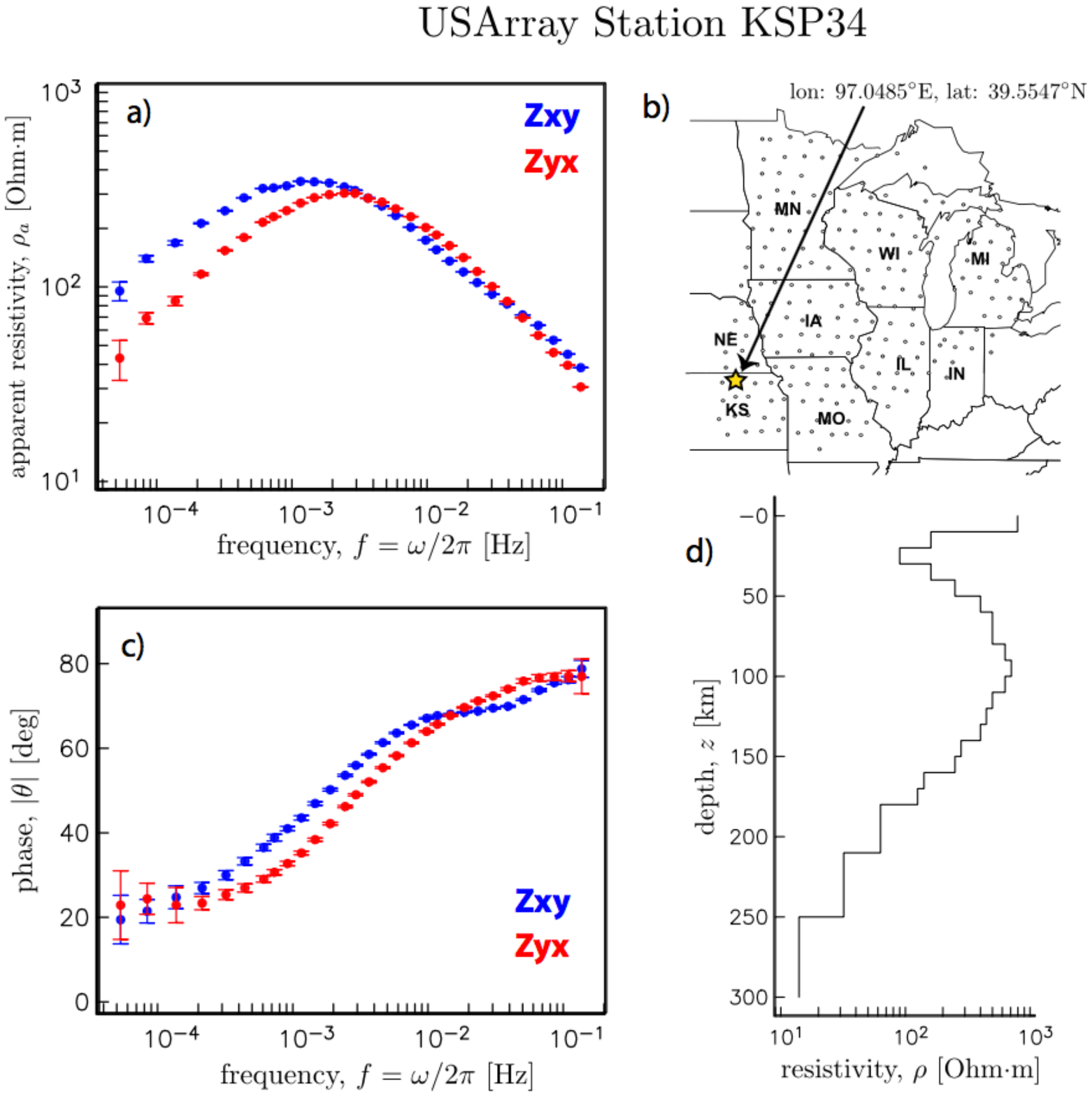}
\caption{Apparent resistivity and phase angle data from USArray MT
  station for KSP34 located NW of Kansas City, KS, USA from the US Array:
a) Apparent resisistivity spectrum based on Zxy (blue) and Zyx (red) elements of the $2\times 2$ impedance tensor $\bold Z$. The
similarity in the curves, especially at high frequencies, is
indicative a locally 1D conductivity profile beneath the observation
point. b) Location map for USArray MT station KSP34. c) Complex phase
angle of the ratios $E_x$/$H_y$ (Zxy, blue) and $E_y$/$H_x$ ($Z_{yx}$, red), again
generally similar and indicating a locally 1D conductivity profile
beneath the station.  d) Depth profile of electrical resistivity
beneath station KSP34 estimated by 3D inversion of all sites in
sub-panel (b)  (figure 3, sub-panel (a), Yang et al., EPSL 2015 ).}
\label{KSP}
\end{figure}


\subsection{Strategies for a spatially--variable fractional exponent}

An initial assumption in problem statement \eqref{cjw1} is the spatial invariance
of the fractional exponent $s$ over the spatial domain $\Omega$. However, if
$s$ is intepreted to represent via non--locality some degree of long--range correlation 
of underlying material properties (e.g. electrical conductivity), then it is relevant 
to consider how spatial variability in this correlation is accommodated in the architecture
of the fractional calculus paradigm.     
In addition, variability in $s$ enables
us to truly capture the non-smooth effects such as fractures by prescribing variable
degree of smoothness across the scales. A detailed analysis for variable $s$, where the
authors have created a time-cylinder based approach, has been recently carried
out in \cite{Antil_and_Rautenberg:2018}. For a precise definition of the fractional
Laplacian with variable $s$ we refer to \cite{Antil_etal:2018b}.

In the case of a piecewise constant $s$, a conceptually simple
strategy is to decompose the domain $\Omega$ into subdomains on which
$s$ is constant and impose our Kato method over each
of the subdomains.  Note that the solution for $w$ in \eqref{cjw3} is
independent of $s$ and may be obtained without any need for domain
decomposition. Although differences in $s$ among domains means that
the number of functions $v_{\ell=0,\ldots,L}$ also varies among
domains, the boundary condition $v=0$ on each of the subdomains
ensures continuity of $v$, and therefore continuity of $u=v+w$
throughout $\Omega$. Observe that computation of $\{v_\ell\}$ in one
subdomain is independent of its calculation in another, and hence,
$\{v_\ell\}$ over each of the subdomains can computed in parallel with
no message passing or interdomain communication required once $w$ is
solved for and shared globally throughout $\Omega$.  That said,
several issues need to be resolved before this idea can be defensibly
implemented.  First, the suggestion of zero (subdomain) boundary
conditions on $\{ v_\ell\}$ needs to be physically justified. If found
to be unsound, the embarrassingly parallel structure just described
will instead require interdomain communication and potentially
interpolation.  Second, because the fractional Laplacian is inherently
non--local, its support extends over the global domain $\Omega$.
Ensuring global extent of non--locality in the context of subdomains
requires further analysis.  Lastly, function and flux continuity for a
given $v_\ell$ within a subdomain is guaranteed; the conditions for
such guarantees, in a general sense, at subdomain boundaries have yet
to be determined.  Because of these complexities, further analysis of
this domain decomposition concept is deferred to future publication.

\subsection{Fractional time derivatives}
Prior work in electromagnetic geophysics in contemplation of Ohm's
constitutive law being represented in terms of fractional calculus
have focused on fractional {\it time} derivatives, rather than the
fractional space derivatives described here
\cite{Weiss_and_Everett:2007,Everett:2009,Ge_etal:2015}.  Such
analyses are comparatively simple in that the fractional space
derivatives $D_z^\alpha$ of (1.2) are replaced by time derivatives
$D_t^\beta$, thus modifying the complex wavenumber as
$k^2 = -(\text{i}\omega)^{1-\beta}\mu_0\sigma$.  Solutions to
\eqref{cjw1} in layered media when $s=1$ (equivalently, $\alpha=0$
since $s=1-{\alpha\over 2}$) follow the usual method of posing
characteristic solutions $\exp(\pm kz)$ in each of the layers,
coefficients for which are determined through enforcement of boundary
condition \eqref{cjw1} along with continuity of $u$ and $\partial_z u$
at layer boundaries.  Solving this time-fractional Helmholtz equation
on the domain $\Omega: z\in[0,z^*=1000]$ m with $u(0) = 1 + \text{i}$,
$u(z^*)=0$ and $\sigma=0.01$ S/m$(\text{rad/s})^{-\beta}$ yields a
characteristic magnetotelluric response (\figref{helm_time}) distinct
from that obtained in the case of space--fractional derivatives
$s\ne 1$ (\figref{sounding}).  As noted in \cite{Ge_etal:2015},
imposing the time--fractional derivative in this way is equivalent to
recasting real--valued electrical conductivity $\sigma$ as a
frequency--dependent, complex--valued conductivity
$\sigma(\text{i}\omega)^\beta$.  The quasi--linear power--law behavior
in apparent resistivity and phase angle (\figref{helm_time}) seen at
high frequencies ($f>100$ Hz) is objectively distinct from that
computed for the space--fractional Helmholtz system
(\figref{sounding}) and offers an unambiguous diagnostic for
discriminating between the two.  These differences have their origin
in the how anomalous power--law diffusion is captured by each. In
the case of fractional time derivatives of order $1-\beta$, as
considered in this latest example, the system is considered
subdiffusive and consistent with an anomalously high likelihood of
long wait times between successive jumps of charge carriers in a
continuous time random walk as might be applied to fluid transport in
a porous medium \cite{Metzler_and_Klafter:2000}.  Instead, the
space--fractional derivatives which occupy the primary focus of the
present study capture long--range interactions (spatial nonlocality)
of charge carriers as a superdiffusive system, perhaps through
inductive coupling (a phenomena absent in the physics of fluid flow in
porous media).  This contrast -- super- versus sub-diffusion -- is the
essence of the causative physics behind the different magnetotelluric
responses predicted by (\figref{sounding} and \figref{helm_time}).

\begin{figure} 
\centering \includegraphics[width=0.75\textwidth]{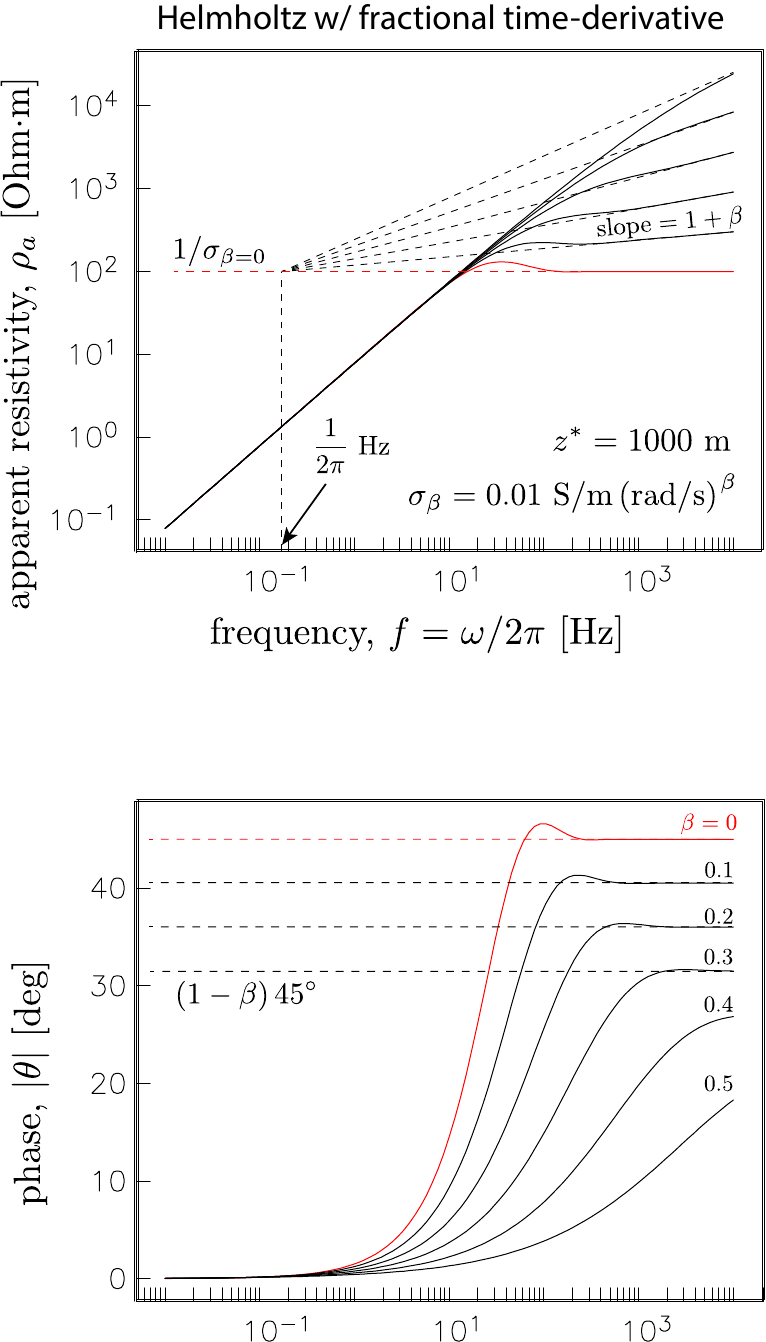}
\caption{Apparent resistivity spectra for classical Helmholtz equation $s=1$ with $\text{i}\kappa^2 = 
(\text{i}\omega)^{1-\beta}\mu_0\sigma$, where the $\beta$ terms arise in the 1D magnetotelluric
case from Ohm's law with a fractional, non--local time dependence attributable to sub-diffusion of 
electric charge following a continuous time random walk with a heavy--tailed distribution of waiting times. 
Compare  to \figref{sounding} for the space--fractional case describing super--diffusion, where the heavy--tailed
distribution of step length (a.k.a.  L\'evy flights) captures long--range interactions between charges.}
\label{helm_time}
\end{figure}

\section{Conclusions}

We have presented a novel, practical solution to the 
fractional Helmholtz equation based on the Kato formulation 
of the fractional Laplacian operator, a lifting (splitting) 
strategy to handle non-zero Dirichlet boundary conditions, 
and finite element discretization of the spatial domain. 
This specific finite element discretization derives from our 
statement of the variational problem, from which alternative
discretizations (orthogonal polynomials, wavelets, 
spectral functions, etc.) offer an interesting direction 
for future research.  
Whereas the analogous Kato/lifting 
strategy for solving the fractional Poisson equation leads to 
decoupled system of integer Laplace solves which can be 
done in parallel with no inter-solve communication, 
solution of the fractional Helmoltz admits
no such decoupling. This leads to a large, block--dense 
system of linear equations upon discretization which significantly
increases the resource requirements for obtaining a numerical
solution. In response, we augment the variational problem by introducing 
an additional unknown which collapses the $L-1$  block coupling 
matrices in a given block--row into a single block matrix, at
the expense of only one additional (dense) block--row in the linear 
system.  For typical problems where with $L >> 100$, the 
added computational burden of this compatibility equation is 
inconsequential, yet the reduction in matrix storage is 
significant, going from $L^2$ to simply $2L$.
Thus, a key feature of this augmented variational
problem is the extreme block--sparsity of the resulting linear 
system of equations, a feature which is independent of the 
choice of discretization and important for efficient solution 
of large--scale systems. Validation of the 
algorithm for linear, nodal finite elements shows 
$h^2$ reduction in RMS error for an MMS test problem -- demonstrating
that our formulation of the fractional Helmholtz problem does 
not corrupt the convergence behavior expected from solution 
of integer-order Helmholtz.

We apply this formulation for fractional Helmholtz to the 
growing body of observational evidence of anomalous diffusion in 
nature -- here, asking the question, ``Does the Earth, with its 
incalculable geologic complexity, respond to electromagnetic
stimulation in a way that is consistent with fractional 
diffusion and the non--locality that is central to the differential
operators of the governing physics''?  Whereas temporal 
non-locality of Maxwell's equations has previously been observed 
as sub--diffusive propagation, the fractional Helmholtz 
equation studied here describes super--diffusion by attributing
fractional derivatives directly to the spatial distribution of 
material properties in Ohm's constitutive law.  Earth electromagnetic
response is computed in the context magnetotelluric (MT) analysis -- 
a classic geophysical exploration technique dating back to
to middle 20th century -- and comparison with the EarthScope
USArray database. We find qualitative agreement between the 
predicted fractional Helmholtz response functions and those
observed at a middle North American measurement site.  This 
congruence in electromagnetic response thereby offers an altenative
interpretation of the MT data at the site, 
one where the classical interpretation of a layered Earth geology 
with deep resistive rocks overlain by a conductive overburden
is contrasted with new interpretation suggesting complex, geologic
texture consistent with the site's proximity significant deep crustal 
tectonic structure.  

Outstanding issues for future research therefore lie in two fundamental
areas: arriving at a clearer mapping between the value of a
fractional exponent $s$ and the material heterogeneity it's intended 
to represent; and, extension of the computational tools to higher
dimension with parallel implementation, including spatially variable and/or
anisotropic $s$ values. The former may be informed, as we've done here, 
by reinterpretation of existing observational data through 
fractional calculus concepts, but augmented by detailed material analysis.
The latter naturally feeds into ongoing
efforts in PDE--constrained optimization for material property 
estimation, now augmented with the desire to recover $s$, too, 
as a measure of material complexity or sub--grid structure. 
Algorithmic advances in multi-level domain decomposition (decomposition 
over physical domain in addition to decomposition over the functional
blocks of global system matrix) will also be required for full
exploration of fractional Helmholtz concepts on large, 3D domains.

\vfill\eject
\bibliographystyle{plain}
\bibliography{lit}
\end{document}